\documentclass[twocolumn,prb,showpacs]{revtex4}
\usepackage{epsfig}
\begin{document}
\topmargin-1.0cm

\title {
Temperature dependent photoluminescence of organic semiconductors
with varying backbone conformation}
\author {S. Guha $^1$, J. D. Rice$^1$, Y. T. Yau $^1$, C. M. Martin$^2$, M.
Chandrasekhar$^2$, H.R. Chandrasekhar$^2$, R. Guentner$^3$, P.
Scandiucci de Freitas$^3$ and U. Scherf$^3$ } \affiliation
{$^1$Department of Physics, Astronomy and Materials Science,
Southwest Missouri State University, Springfield MO 65804 USA}
\email[E-mail: ]{sug100f@smsu.edu} \affiliation {$^2$Department of
Physics, University of Missouri, Columbia, MO 65211 USA}
\affiliation {$^3$Institut f\"ur Chemie and Polymerchemie,
Universit\"at Potsdam, Germany}
\date{\today}

\begin{abstract}
We present photoluminescence studies as a function of temperature
from a series of conjugated polymers and a conjugated molecule
with distinctly different backbone conformations. The organic
materials investigated here are: planar methylated ladder type
poly {\em para}-phenylene, semi-planar polyfluorene, and
non-planar {\em para} hexaphenyl. In the longer-chain polymers the
photoluminescence transition energies blue shift with increasing
temperatures. The conjugated molecules, on the other hand, red
shift their transition energies with increasing temperatures.
Empirical models that explain the temperature dependence of the
band gap energies in inorganic semiconductors can be extended to
explain the temperature dependence of the transition energies in
conjugated molecules.

\end{abstract}

\maketitle

\section{Introduction} \label{sec:intro}

Conjugated organic molecules such as short-chain oligomers and
longer-chain polymers are very promising active materials for
low-cost, large-area optoelectronic and photonic devices.\cite{1}
Semiconducting properties are defined by the ability of these
materials to efficiently transport charge (holes or electrons)
along the chain due to their $\pi$-conjugation or between adjacent
chains due to the $\pi$-orbital overlap of neighboring molecules.
Devices such as organic light-emitting diodes (OLEDs),
transistors, and photodiodes are currently attracting much
attention.  Blue electroluminescent materials are of particular
interest for organic displays since blue light can easily be
converted into red and green by color-changing media (fluorescent
dyes).  Commercial availability of organic light-emitting diodes
has attracted growing attention to $\pi$-conjugated molecules such
as pentacene, oligothiophene, oligomers of poly {\em
para}-phenylene (PPP) and polymers\cite{5,6} such as poly {\em
para}-phenylene vinylene(PPV), polythiophene and PPP. In recent
years polyfluorenes (PF) have emerged as attractive alternatives,
showing the highest photoluminescence quantum efficiency (55\%)
compared to other conjugated polymers/molecules in solid
state\cite{7} and also maintain a high hole mobility at room
temperature.\cite{8}  Since high photoluminescence quantum yield
(PLQY) is the primary consideration in devices such as OLEDs,
mechanisms that change the PLQY are crucial to the understanding
and design of these devices. Optical processes that reduce the
PLQY include competing non-radiative processes, transitions that
depopulate the ground state due to reabsorption of photons from
singlet states, or situations that can involve intermolecular
interactions.

The temperature dependence of electronic states in inorganic
semiconductors and heterostructures has been studied extensively
in the last 30 years or so. In these systems the band gaps exhibit
large shifts and lifetime induced broadenings as a function of
temperature.\cite{adachi} One of the first empirical models to
explain the temperature dependence of the band gap energy in bulk
semiconductors was put forward by Varshni.\cite{9}  Subsequently,
there have been other empirical models taking into account an
average Bose-Einstein statistical factor for phonons to explain
the temperature dependence of the electronic states.\cite{10,11}
This latter model fits the temperature dependence of the
transition energies better in narrow-well inorganic semiconductor
heterostructures and superlattices.\cite{12}

In inorganic semiconductors the two mechanisms that are
responsible for the temperature dependence of energy bands at
constant pressure are thermal expansion and renormalization of
band energies by electron-phonon interactions. Typically the
former mechanism has a negligible effect so that the changes in
the band gap energies arise primarily from electron-phonon
interactions. Theoretical calculations of the temperature
dependence of band energies in Si and Ge that take into account
such electron-phonon interactions agree very well with the
experimental results.\cite{13}

The motivation of this work is to understand the temperature
dependence of the electronic states in organic semiconductors
using steady state photoluminescence (PL) spectroscopy and to test
the extension of the existing models for inorganic semiconductors
to organic materials. We compare different conjugated materials,
namely long-chain polymers and short-chain oligomers with distinct
conformational variations in their backbone.

Spectroscopic properties of conjugated molecules and polymers are
highly sensitive to changes in backbone conformation. The three
families of conjugated materials that we compare in this work have
distinct differences in their backbone conformation. They include
planar methyl substituted ladder-type poly {\em para}-phenylene
(MeLPPP), semi-planar poly[2,7-(9,9'-bis(2-ethylhexyl]fluorene
(PF2/6) and non-planar ({\em para} hexaphenyl) (PHP). All three
materials are of technological importance due to their strong blue
luminescence and high chemical purity. They have been used as
active materials in OLEDs.\cite{14,15,7} Both MeLPPP and PF2/6 are
long-chain processable conjugated polymers\cite{setayesh} whereas
PHP is a short-chain oligophenyl that forms monoclinic
crystallites of space group P2$_1$/a.\cite{16}

\begin{figure}
\unitlength1cm
\begin{picture}(5.0,8.0)
\put(-2.5,0.9){ \epsfig{file=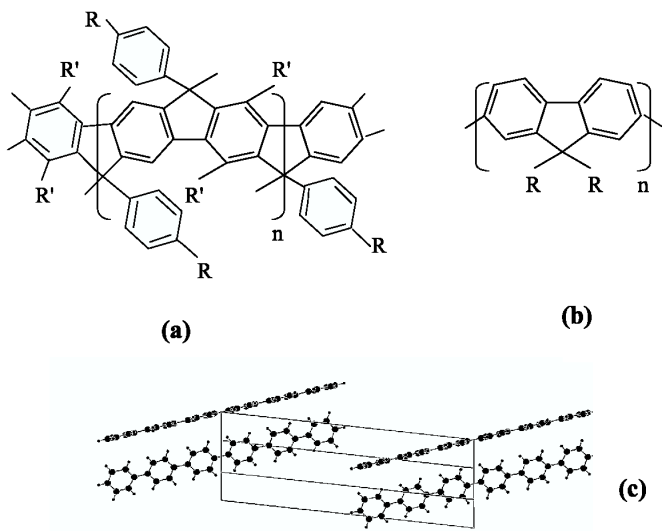, angle=0, width=8.8cm,
totalheight=7.0cm}}

\end{picture}
\caption{Chemical structure of  (a) MeLPPP (b) PF2/6  and (c) PHP.
In (a) R and R' refer to C$_{10}$H$_{21}$ and C$_6$H$_{13}$,
respectively and in (b) R refers to the ethyl hexyl side chain.}
\label{figure1}
\end{figure}

PHP is characterized by a torsional degree of freedom between
neighboring phenyl rings. In the crystalline state the molecules
are arranged in layers, with a herringbone type arrangement found
in each layer as shown in Fig. \ref{figure1}(c). PHP can be
planarized by the application of hydrostatic pressure.\cite{17} On
the average, at room temperature it is more planar than at lower
temperatures. MeLPPP is amorphous due to the bulky side groups.
Neighboring phenyl rings are planar due to the methyl bridges and
show no torsional degree of freedom (see Fig. \ref{figure1}(a)).
The planarity between phenyl rings results in a high intrachain
order and a low defect concentration. This is attributed to the
synthesis method, which is highly selective in forming only
certain bonds, and hence reduces the number of defects.\cite{18}

PF2/6, on the other hand, can be viewed as a semi-planar polymer.
It forms planar monomer units but has a torsional degree of
freedom between adjacent monomer units as shown in Fig.
\ref{figure1}(b). The alkyl side chains of the fluorene moieties
have been shown to strongly influence the solid-state packing of
the molecule.\cite{19} We also compare the spectra of two other
derivatives of PF: a dialkylated copolymer,
poly[9,9-bis(3,7,11-trimethyldodecyl)fluorene-2,7-diyl] (PF1112)
and poly[9,9-bis(4-(2-ethylhexoxy)phenyl) fluorene] (PF-P). The
copolymer has longer alkyl side chains compared to PF2/6 with 2\%
of 2,7-fluorenone units incorporated in the backbone as a model
for a photodegeneration-induced defect-rich polyfluorene. PF-P is
a diphenyl-substituted PF exhibiting an extraordinarily small
defect concentration.

Upon further processing some polyfluorene films display a $\beta$
phase, which has a more extended intrachain $\pi$-conjugation, in
addition to the regular glassy $\alpha$ phase. The $\beta$ phase
has been detected in 9,9-di-n-octyl-PF  (PF8 or PFO) upon thermal
cycling from 80 to 300 K\cite{20} and slowly reheating it to room
temperature or exposing a film to the vapor of a solvent.\cite{21}
The $\beta$ phase shows a distinct red shift of absorption and
emission peaks with a well-resolved vibronic progression both in
absorption and emission. In contrast, the $\alpha$ phase shows a
well-resolved vibronic progression only in the emission spectrum.
Using x-ray and electron diffraction measurements, Lieser
\textit{et al}. have shown that the $\beta$ phase is completely
absent in PF with branched side groups like PF2/6.\cite{22}

This paper is organized as follows; Section II describes the
experimental setup. In Section III, we present the PL results from
MeLPPP, PF, and PHP as a function of temperature. Section IV is a
discussion of our results, followed by our conclusions in Section
V.

\section{Experimental Details}\label{sec:exptaldetails}

Highly purified PHP powder was obtained from Tokyo Chemical
Industries Ltd. PL spectra were measured from films of MeLPPP, PF,
and PHP. The MeLPPP and PF2/6 films were prepared by spin coating
on a glass slide from a toluene solution, and their thickness was
~ 1000 \AA . We also prepared a thicker film (~30,000 \AA ) by
dropcasting PF2/6. In this paper PF(A) and PF(B) refer to the
thicker and thinner PF2/6 films, respectively. The films were
dried at room temperature. The PHP film was prepared by vacuum
evaporation. PHP was evaporated in high vacuum (~10$^{-6 }$ mbar)
on regular glass substrates. PHP powder was put in a quartz
crucible that was heated by passing 73 A of current. The thickness
of the sample was ~1000 \AA . The PL spectra were excited using
the 363.8 nm line of an $\rm Ar^+$ laser. The luminescence
excitation was analyzed with a SPEX 0.85 m double monochromator
equipped with a cooled GaAs photomultiplier tube and standard
photon counting electronics. The samples were loaded in a cryostat
and evacuated to below 100 mTorr to prevent photo-oxidative
damage. For low temperature measurements a closed cycle
refrigerator was employed.

\section{PL RESULTS}\label{sec:PLresults}

A vibronic progression is seen in the PL emission for all the
materials under consideration, indicating a coupling of the
backbone carbon-carbon stretch vibration to electronic
transitions. The vibronic spacing in all the samples lies between
1300 and 1400 $\rm cm^{-1}$. The vibronic peaks result from a
non-zero overlap of different vibronic wavefunctions of the
electronic ground and excited states. The emissive transition
highest in energy is called the 0-0 transition, which takes place
between the zeroth vibronic level in the excited state and the
zeroth vibronic level in the ground state. The 0-1 transition
involves the creation of one phonon. In the adiabatic picture,
vibronic progression in the electronic spectra implies that the
ground and excited state equilibrium structures are displaced
relative to one another in configuration space. The spectral
intensity is approximated by the superposition of transitions
between the vibrational frequencies of the ground- and the excited
electronic states. In the emission process the probability of the
0$^{th}$ vibronic excited state to the n$^{th}$ vibronic ground
state is given by

\begin{equation}\label{1}
I_{0\to n}=\frac{e^{-S}S{^n}}{n!}
\end{equation}
where $S$, the Huang-Rhys factor, is given
by$M\omega/2\hbar({\Delta})^2$.\cite{23} Here $\omega$ is the
vibrational frequency, $M$ is the reduced mass of the harmonic
oscillator that couples to the electronic transition and $\Delta$
is the displacement of the potential curve between the ground and
excited electronic states. The Huang-Rhys factor therefore
corresponds to the average number of phonons that are involved
when the excited molecule relaxes from its ground state
configuration to the new equilibrium configuration in the excited
state (after the absorption of a photon) and $S\hbar\omega$ is the
relaxation energy.  If we assume that $\omega$ is the same for
ground and excited states and that the potentials are perfectly
parabolic, $S$ can be determined from the fractional intensity of
the vibronic peaks. In particular,
\begin{equation}\label{2}
S=(I_{0\to 1}+2I_{0\to 2}+3I_{0\to 3})/I_{total}
\end{equation}
$I_{0\to 1}$, $I_{0\to 2}$, and $I_{0\to 3}$ refer to the
intensity of the emission from the zeroth vibrational level
excited state to the first, second and third vibrational level of
the ground state, respectively. $I_{total}$ is the total intensity
of the individual vibronics. Here we assume that the transition
matrix elements are the same for all vibronics and neglect all
vibronics above 0-3.

\subsection{MeLPPP Film}
\begin{figure}
\unitlength1cm
\begin{picture}(5.0,6.0)
\put(-2.5,-0.7){ \epsfig{file=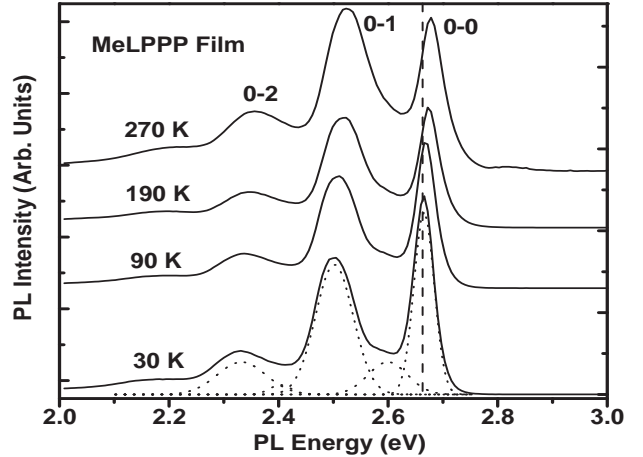, angle=0, width=9.5cm,
totalheight=7.5cm}}

\end{picture}
\caption{PL spectra of a MeLPPP film at a few values of
temperature. The dotted line under the 30 K data shows the actual
fit to the data. The vertical dashed line indicates a blue shift
of the 0-0 vibronic peak with increasing temperature.}
\label{figure2}
\end{figure}

Figure \ref{figure2} shows the PL spectra of an MeLPPP film at a
few selected values of temperature. The PL spectrum was fitted
with Gaussians in order to obtain the individual peak positions,
amplitudes and broadening (FWHM) parameters. The three main
vibronic peaks observed are labeled as the 0-0, 0-1 and the 0-2.
There are additional vibronic peaks observed between the main
vibronic peaks that are also seen in other works.\cite{24}  The
dotted line under the 30 K spectrum is a representative of our
fits. At 30 K the main vibronic peaks that are observed are the
0-0 peak at 2.67 eV, the 0-1 at 2.5 eV, 0-2 vibronic peak at 2.33
eV, and the 0-3 peak at 2.19 eV. In addition, vibronic replicas
are observed at 2.61 eV and 2.4 eV. The energy difference between
the main successive vibronic peaks is ~0.17 eV indicative of the
coupling of the =C-C=C-C= stretch vibration to the conjugated
backbone. At all temperatures the PL spectra were fit with the
same number of Gaussian peaks for consistency. The overall
spectrum not only blue shifts with increasing temperatures but the
relative intensities of the individual vibronic peaks change as
well.

\begin{figure}
\unitlength1cm
\begin{picture}(5.0,6.)
\put(-3,-0.3){ \epsfig{file=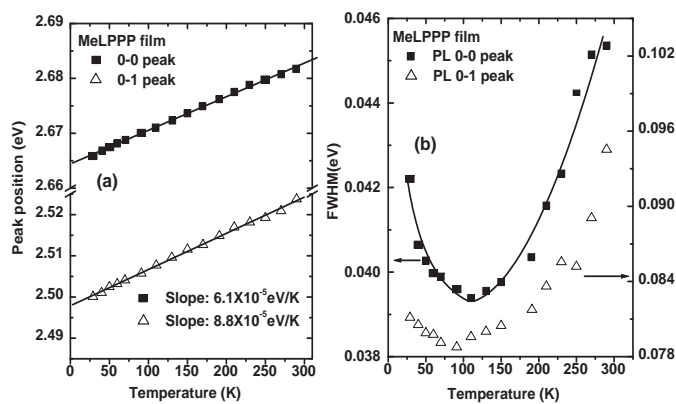, angle=0, width=10.5cm,
totalheight=7.cm}}
\end{picture}
\caption{(a) The peak position of the 0-0 and the 0-1 PL
transitions as a function of temperature in MeLPPP (b) FWHM of the
0-0 and the 0-1 peak as a function of temperature in MeLPPP.}
\label{figure3}
\end{figure}

Figure \ref{figure3} (a) shows the energy position of the 0-0 and
the 0-1 vibronics as a function of temperature. The 0-2 peak in
MeLPPP also shows a similar behavior. The average value of the
rate of shift is 7.5$\times$10$^{-5}$ eV/K. Figure \ref{figure3}
(b) shows the FWHM (eV) for the 0-0 and the 0-1 vibronic peaks as
a function of temperature. The bold line is a guide to the eye.
The line widths decrease till ~100 K and then increase beyond 100
K. This cannot be attributed to any fitting artifact: at all
temperatures the PL spectrum is fitted with same number of
individual vibronic peaks (shown in the 30 K data of Fig.
\ref{figure2}) which are allowed to vary in position, amplitude
and width till the best fit is obtained. Moreover, both the 0-0
and the 0-1 vibronic peaks show the same trend.

\subsection{PF Film}

\begin{figure}
\unitlength1cm
\begin{picture}(5.0,6.)
\put(-2.,-0.5){ \epsfig{file=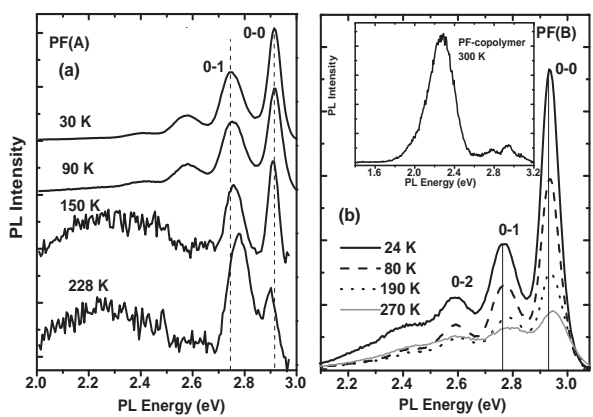, angle=0, width=9.2cm,
totalheight=7.3cm}}
\end{picture}
\caption{PL spectrum of PF2/6 at selected values of temperature
for a  (a) thicker and  (b) thinner film. The vertical lines show
the shift in the transition energies with temperature. The inset
in (b) is the PL spectrum of PF-copolymer at 300 K. The strong 2.3
eV peak is related to the emission from the keto defects.}
\label{figure4}
\end{figure}

\begin{figure}
\unitlength1cm
\begin{picture}(5.0,6.)
\put(-2.5,-0.45){ \epsfig{file=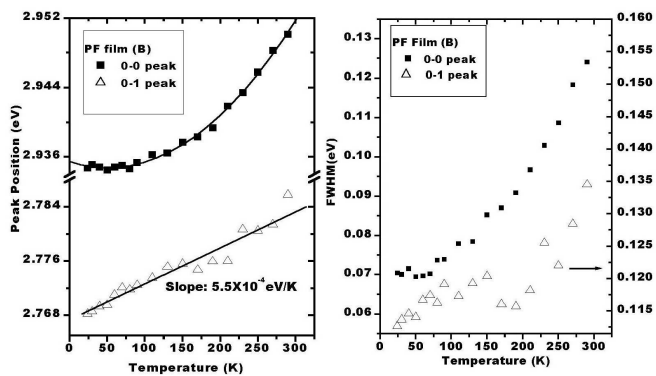, angle=0, width=9.cm,
totalheight=7.0cm}}
\end{picture}
\caption{(a) Peak position of the 0-0 and the 0-1 peak in PF(B) as
a function of temperature. (b) FWHM of the 0-0 and the 0-1 peak in
PF(B) as a function of temperature.} \label{figure5}
\end{figure}

Figure \ref{figure4} shows the PL spectra from two PF2/6 films
(with different thickness) for a few selected values of
temperature. The relative intensity of the 0-0 peak to the 0-1
peak in PF(A) (thicker film) is lower compared to PF(B) (thinner
film) indicating a higher self-absorption in PF(A). At 30 K the
main vibronic peaks that are observed are the 0-0 peak at 2.93 eV,
the 0-1 at 2.77 eV and the 0-2 transition at 2.59 eV in PF(B). An
additional vibronic replica is observed at 2.86 eV between the 0-1
and 0-2 peaks. The PL transition energies show a blue shift with
increasing temperatures, similar to the MeLPPP sample. The peak
positions and the FWHM of the 0-0 and the 0-1 PL vibronics are
shown in Fig. \ref{figure5}. The sublinear behavior of the 0-0
peak position in PF(B) with increasing temperatures may be related
to self-absorption effects.

The PL spectra of PF(A) show  a slightly different behavior as
seen in Fig. \ref{figure4}(a). The 0-0 peak red shifts whereas the
0-1 peak blue shifts with increasing temperatures. The red shift
of the 0-0 peak in this sample is most probably an artifact due to
self-absorption effects. With increasing temperatures a broad peak
at 2.3 eV emerges at around 150 K, shown in Fig. \ref{figure4}(a).
Recent work suggests that this peak is related to the emission
from keto defects sites (9-fluorenone sites). \cite{19,25}  These
defect sites act as guest emitters which can efficiently trap
singlet excitons created on the conjugated polyfluorene backbone
by a dipole-dipole induced F\"{o}rster-type energy transfer.
\cite{26} The keto defect sites can be accidentally incorporated
into the $\pi$-conjugated PF backbone due to the presence of
non-alkylated or monosubstituted fluorene sites during synthesis
or as a result of a photo-oxidative degradation process. In order
to reduce photodegradation due to the exciting UV, the samples
were kept in vacuum during our PL measurements. The concentration
of these defect sites is quite low in the PF2/6 sample since the
2.3 eV emission is absent in the thinner PF(B) film. Also, it is a
thermally activated process; in PF(A) the defect related emission
is only observed for temperatures above 150 K. The inset of Fig.
\ref{figure4} (b) shows the PL spectrum from PF1112, the copolymer
with 2{\%} incorporated fluorenone sites. The strong 2.3 eV
emission is from the keto defects which overwhelms the emission
from the PF backbone.

\begin{figure}
\unitlength1cm
\begin{picture}(5.0,7.0)
\put(-2.,0.3){ \epsfig{file=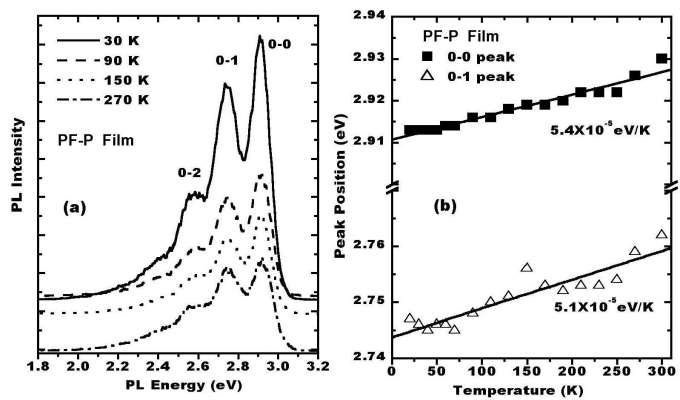, angle=0, width=9.3cm,
totalheight=7.0cm}}

\end{picture}
\caption{(a) PL spectra of PF-P at a few values of temperature.
(b) The 0-0 and the 0-1 peak position as a function of temperature
in PF-P.} \label{figure6}
\end{figure}

In contrast, the diphenyl-substituted PF (PF-P) is expected to
have almost no keto defects due to a different synthesis of the
corresponding monomer blocks that prevents non- or monosubstituted
fluorene sites. We compare the temperature dependent PL from a
drop-casted film of PF-P to the PF2/6 sample of film thickness
comparable to the sample PF(A). Figure \ref{figure6} (a) shows the
PL spectra from PF-P for selected values of temperature. Clearly
the 2.3 eV emission is not observed at higher temperatures unlike
in PF(A), indicating that the sample is almost free of keto defect
sites. Figure \ref{figure6} (b) shows the peak position of the 0-0
and the 0-1 vibronics as a function of temperature. The average
value for the rate of shift of the PL vibronics is
5.2$\times$10$^{-5}$ eV/K.

\subsection{PHP}

\begin{figure}
\unitlength1cm
\begin{picture}(5.0,7.0)
\put(-2.,0.3){ \epsfig{file=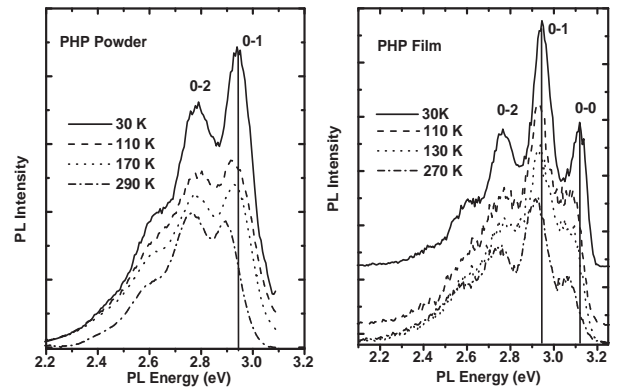, angle=0, width=9.8cm,
totalheight=7.0cm}}
\end{picture}
\caption{PL spectra at four selected values of temperature for (a)
PHP powder and (b) a thin evaporated PHP film.} \label{figure7}
\end{figure}

\begin{figure}
\unitlength1cm
\begin{picture}(5.0,6.0)
\put(-2.,-1.3){ \epsfig{file=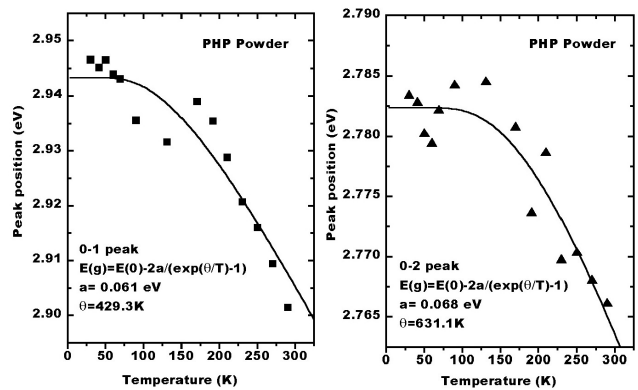, angle=0, width=10.0cm,
totalheight=8.0cm}}

\end{picture}
\caption{Peak position of the (a) 0-1 and the (b) 0-2 PL
transition energies from PHP powder as a function of temperature.
The bold line is a fit to Eq. (3).} \label{figure8}
\end{figure}

The PL measurements from PHP were measured both from the powder
and from a thin evaporated film. Figure \ref{figure7} shows the PL
spectra both from the powder and film for selected temperatures.
The 0-0 transition is not observed in the powder due to
self-absorption. The film also shows a certain amount of
self-absorption since the relative intensity of the 0-0 peak is
smaller compared to the 0-1 or the 0-2 transition peak. At 30 K
the main vibronic peaks that are observed are the 0-0 peak at 3.12
eV, the 0-1 at 2.95 eV and the 0-2 vibronic peak at 2.78 eV.
Additional vibronic peaks are observed at 3.07 eV and 2.86 eV. The
individual vibronics red shift with increasing temperatures both
in the film and powder. A red shift of the PL spectrum with
increasing temperatures in PHP has been observed before, but a
detailed analysis was not carried out.\cite{17,27} The PHP film
was more prone to local heating effects during the PL measurement;
we therefore restrict our analysis to the PHP powder. Figures
\ref{figure8}(a) and (b) show the individual positions of the 0-1
and the 0-2 peaks, respectively, from the PHP powder. The bold
line is a fit to an empirical model, discussed in detail in
Section \ref{sec:discussion}.

\section{DISCUSSION}\label{sec:discussion}

Conjugated polymers can be viewed as an inhomogeneous collection
of varying-length chain segments. Time- and frequency-resolved
optical studies show that after a photon excites a chain segment
with energy above the threshold, the exciton that is created
executes a random walk through the density of states via
F{\"o}rster energy transfer,\cite{26} until it becomes trapped on
a low-energy site, from where emission occurs. These sites
presumably have a higher conjugation length.\cite{28,29}

The temperature dependence of the PL energies is different for
long-chain conjugated polymers compared to the shorter chain
oligomers; conjugated polymers show a blue shift of PL energies
with increasing temperatures, whereas conjugated molecules like
PHP show a red shift with increasing temperatures. It is
interesting to point out that although the two types of polymers
studied here have differences in their backbone conformation,
MeLPPP being planar and PF2/6 is semi-planar, both show the same
trend as a function of temperature. A similar behavior has been
observed in other conjugated polymers such as PPV\cite{28} and
MEH-PPV.\cite{30,31} Though the general trend in PL energies as a
function of temperature is similar in the conjugated polymers
investigated in this work, there are differences in the magnitude
of their shifts with increasing temperatures.

\begin{figure}
\unitlength1cm
\begin{picture}(5.0,7.5)
\put(-1.5,-0.7){ \epsfig{file=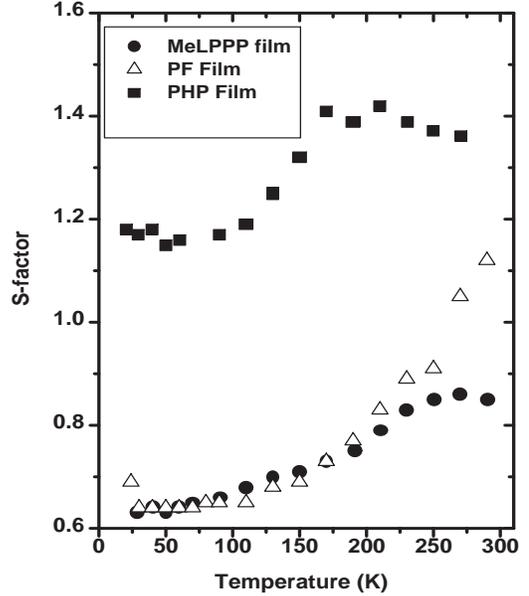, angle=0, width=8.0cm,
totalheight=9.0cm}}

\end{picture}
\caption{Huang-Rhys factor versus temperature for MeLPPP, PF and
PHP films. $S$ was calculated using Eq. {\ref{2}}} \label{figure9}
\end{figure}

The relative strengths of the vibronic transitions also change
along with shifts in energy with temperature. The Huang-Rhys
factor decreases both for the polymers and PHP with decreasing
temperatures, as shown in Fig. \ref{figure9}. The $S$-factor was
calculated using Eq. (\ref{2}). Smaller molecules exhibit a larger
$S$-factor due to their large normal coordinate displacements. For
a detailed explanation see Hagler \textit{et al}. \cite {30}. It
is therefore not surprising that PHP has a higher $S$-factor
compared to the polymers. From Fig. \ref{figure9} it is seen that
PHP shows a saturation effect at higher temperatures. This may be
related to the fact that since the electronic transition energies
red shift with increasing temperatures, there may be a higher
overlap of the 0-0 peak with absorption resulting in a further
decrease in the intensity of the 0-0 peak in addition to the
effect of temperature. A decrease in $S$ with decreasing
temperatures has been observed in other works and is interpreted
as an effect arising from increased conjugation due to exciton
delocalization.\cite{28} This picture by itself cannot explain the
temperature dependence of the PL from both the conjugated polymers
and the shorter molecules, since PHP clearly shows a blue shift of
transition energies with decreasing temperatures. We also observe
the PL emission from quaterphenyl powder to blue shift with
decreasing temperatures. Therefore we must really look at these
systems separately: the conjugated polymers that have a
distribution of chain lengths and the shorter molecules that have
more or less the same chain length distribution.

The electronic energies in bulk inorganic semiconductors display
temperature dependence mainly due to renormalization of band
energies by electron-phonon interactions. The temperature
dependence of the interband transitions can be described with an
expression in which the energy thresholds decrease proportional to
the Bose-Einstein statistical factors for phonon emission plus
absorption\cite{10}
\begin{equation}\label{3}
E_{g}(T)=E(0)-\frac{2a}{[\exp(\Theta/T)-1]}
\end{equation}
where $E(0)$ is the band gap energy at 0 K, $a$ is the strength of
the exciton-phonon interaction. This includes contribution both
from the acoustical and optical phonons. $\Theta$ is the average
phonon temperature. By fitting the 0-1 and the 0-2 PL vibronics of
PHP with Eq. {\ref{3}} (see Fig. \ref{figure8}), we obtain an
average value of the strength of the exciton-phonon interaction as
0.065 eV and the average phonon temperature to be ~530 K.  These
values are comparable to other inorganic semiconductors, for
example in bulk GaAs, $a$=0.057 eV and $\Theta$ =240 K. The PL
linewidth of the individual vibronics in PHP show a scatter with
increasing temperature and on the average remains a constant.

Although the polymers, MeLPPP, PF2/6 and PF-P all show a blue
shift of their PL energies with increasing temperatures, they
shift at different rates. The PL vibronics in MeLPPP and PF-P
shift at rates of 7$\times$10$^{-5}$ eV/K and 5$\times$10$^{-5}$
eV/K, respectively (Figs. \ref{figure3}(a) and \ref{figure6}(b)).
PF2/6, on the other hand, shows a higher shift of the 0-1 PL
vibronic peak (~55$\times$10$^{-5}$ eV/K) as shown in Fig.
\ref{figure5}(a). The different rates have to do with the
differences in their conjugation length, especially since their
molecular weights are not that different. PF2/6 is a more flexible
polymer due to the alkyl substitution unlike MeLPPP and PF-P that
have aryl substituents. PF2/6 undergoes some additional geometry
changes (higher disorder in the chain) with increasing
temperature; the higher rate of the temperature dependence of the
PL transitions may be related to these additional disorder
processes in the polymer. PF2/6 and PF-P show a broadening of the
PL line widths with increasing temperatures. Beyond 100 K, MeLPPP
shows a broadening of the PL line widths as seen in Fig.
\ref{figure3} (b).

A red shift of the PL energies with decreasing temperatures from
the two families of the conjugated polymers with planar backbone
confirmation (MeLPPP) and semi-planar backbone confirmation (PF2/6
and PF-P) confirms B\"assler and Schweitzer's\cite{23} argument
that the process should not really depend upon the freezing out of
torsional modes. Otherwise, MeLPPP, which has a planar backbone
conformation, should show a different behavior. The shifts in the
PL energies reflect more on the temperature dependence of the
relaxation process. A plausible explanation is that upon
increasing the temperature, the excitons that are created on the
polymer backbone do not easily migrate to the low energy segments;
they remain localized on the shorter chain segments that have
higher energies.

In PHP the main contribution to the temperature dependence of
electronic energy arises from the exciton-phonon interaction term,
similar to that in inorganic semiconductors. A possible scenario
is that in the longer chain polymers both renormalization of band
energies due to electron-phonon interaction as well as prevention
of energy migration to lower energy sites play a role, and the
latter appears to be dominant.

It is interesting to point out that in inorganic semiconductors
such as GaN quantum dots (QDs), PL energies blue shift with
increasing temperatures, similar to the conjugated polymers.
\cite{32} This is in contrast to bulk GaN,\cite{33} where the band
energies red shift with increasing temperatures. In the quantum
dot system since there is a distribution of the size of the dots;
increasing temperatures result in a preferential loss of carriers
from larger QDs (which have lower energies), resulting in a net
blue shift of the transition energy due to the emission from the
smaller dots.

\section{CONCLUSION} \label{sec:conclusion}

We have presented a systematic temperature dependent steady-state
PL studies from a series of conjugated polymers and a molecule
with variations in their backbone conformations. The conjugated
polymers show a blue shift of their PL transition energies with
increasing temperature, independent of their actual backbone
conformation. MeLPPP, which is planar and PF that has a
semi-planar backbone confirmation show a similar trend in their PL
energies as a function of temperature. The shifts in the
electronic energies reflect on the temperature dependence of the
actual relaxation process whereby the migration of excitons to the
longer chain segments is hampered with increasing temperatures.

The shorter conjugated molecule, PHP, which has a similar
distribution of chain lengths, shows the opposite trend: a red
shift of the transition energies with increasing temperature,
indicating a renormalization of band energies due to
electron-phonon interaction. This  behavior is similar to that
observed in bulk inorganic semiconductors and can be described by
an empirical model that takes into account the Bose-Einstein
statistical factors for phonon emission and absorption.

\begin{acknowledgments}
Work at SMSU was partly funded by an award from the Research
Corporation {\#}CC5332 and the Petroleum Research Fund
{\#}35735-GB5. S.G. would also like to acknowledge the SMSU summer
faculty fellowship. U.S. thanks SONY International Europe,
Stuttgart, and the Deutsche Forschungsgemeinschaft (DFG) for
financial support.
\end{acknowledgments}



\end{document}